\documentclass[aps,prb,preprint,groupedaddress,floats,tightenlines,superbib]{revtex4}
\usepackage{amssymb}

\usepackage{graphics}
\usepackage{graphicx}
\usepackage{amsmath}

\begin{document}

\title{Toy models for the falling chimney}
\author{Gabriele Varieschi}
\email{gvarieschi@lmu.edu}
\homepage{http://myweb.lmu.edu/gvarieschi/}
\author{Kaoru Kamiya}
\email{kkamiya@lion.lmu.edu}
\affiliation{Department of Physics, Loyola Marymount University, One LMU Drive, Los
Angeles, CA 90045}
\date{\today }

\begin{abstract}
In this paper we review the theory of the ``falling chimney'', which deals
with the breaking in mid-air of tall structures, when they fall to the
ground. We show that these ruptures can be caused by either shear forces,
typically developing near the base, or by the bending of the structure,
which is caused primarily by the internal bending moment.

In the latter case the breaking is more likely to occur between one third
and one half of the height of the chimney. Small scale toy models are used
to reproduce the dynamics of the falling chimney. By examining photos taken
during the fall of these models we test the adequacy of the outlined theory.
This type of experiment, easy to perform and conceptually challenging, can
become part of a rotational mechanics lab for undergraduate students.
\end{abstract}

\maketitle

\section{Introduction}

\label{intro}

One of the most interesting demonstrations for an introductory mechanics
course is the ``Falling Chimney - Free Fall Paradox,'' as it was named by
Sutton in his classical book \textit{Demonstration Experiments in Physics.}%
\cite{Sutton3} In the original version of this demonstration a ball is
placed at one end of a uniform stick, which is pivoted at the other end and
makes initially an angle of about $30^{\circ }$ with the horizontal. The
elevated end of the stick is suddenly dropped, together with the ball, thus
showing a very counter-intuitive behavior. The falling end of the stick
accelerates at a greater rate than the free-falling ball, proving that its
acceleration is greater than $g$, the acceleration of gravity.

A simplified version of the experiment can be performed just with a meter
stick and a coin. The stick is supported in the horizontal position by two
fingers, placed near the two ends. A coin is set on the stick near one end,
which is suddenly released. The effect is similar to the previous
demonstration: the falling end of the rotating stick eventually acquires an
acceleration greater than that of the freely falling coin, which loses
contact with the stick surface and lags behind the falling stick.

A description of the first version of the experiment can be found in almost
every book of physics demonstrations,\cite{Hilton2,Freier,DickRae} sometimes
with different names (``Falling Stick'', ``Hinged Stick and Falling Ball'',
and others). Photographic descriptions or even video-clips of this demo can
be found on-line in several web-pages (see our web-page,\cite{kaoru} for a
collection of related links).

In addition, countless papers exist in the literature; we have traced
several of these, from the 1930's to the present. Some of the earliest
discussions can be found in Constantinides\cite{Constantinides} and Ludeke%
\cite{Ludeke} (as well in the book by Sutton\cite{Sutton3}), followed by
many others.\cite{Hilton1,Young,Phelps,Theron,Hartel} These concentrate
mostly on the simple explanation of the effect, which relies on the concept
of ``center of percussion'' of the rotating stick (a simple introduction to
this concept can be found in Bloomfield\cite{Bloomfield}). This particular
point of the stick (located at a distance from the hinged end equal to two
thirds of the length, for a uniform stick) is moving with the same
acceleration as a particle under gravity, constrained to move along the same
circular path. Points on the stick beyond the center of percussion descend
with accelerations greater than that of particles freely moving under
gravity, on their respective circular paths. As a consequence of this, if
the initial angle formed by the stick with the horizontal is less than about 
$35^{\circ }$, the end point will possess at all times a vertical component
of the acceleration greater than $g$, producing the effect described above.

Several variations of the basic demonstration also exist, \cite%
{Sutton2,Miller,Ficken,Bartlett1,Adams,Haber} the majority of which suggest
attaching an additional mass to the rotating stick at different positions.
The effect for the student or the viewer is even less intuitive than the
original version: an additional mass placed near the end of the stick
actually reduces the acceleration of the end point, affecting substantially
the outcome of the experiment. In general the addition of a mass at any
point on the stick will increase both the total torque on the system (thus
increasing the rotational acceleration) and the moment of inertia of the
system around the axis of rotation (resulting in a decreased rotational
acceleration). The center of percussion of the stick still plays a key role:
if the additional mass is placed beyond it, the effect of the increased
moment of inertia dominates and the acceleration of the rotational motion
will be reduced. If the mass is placed before the center of percussion, the
increase in the torque will dominate and the rotational motion will be
enhanced. The effect is null if the mass is placed exactly at the center of
percussion (a complete discussion of this effect can be found in Bartlett%
\cite{Bartlett1} and Haber-Schaim\cite{Haber}).

The next logical step is to analyze the behavior of a \textit{real falling
chimney}. Almost invariably a tall chimney, falling to the ground like the
stick in the previous discussion, will break in mid-air at some
characteristic height. This is well documented in several photos reproduced
in the literature, such as the one which appeared on the cover of the
September 1976 issue of The Physics Teacher (other photos can be found in
Bundy\cite{Bundy} and Bartlett,\cite{Bartlett2} or also on our web-page\cite%
{kaoru}). The causes of such breaking, the height of the rupture point and
the angle at which the breaking is most likely to occur, are the most
natural questions which arise.

The first analysis\cite{Sutton1} of this problem compared the fall of the
real chimney to the fall of the hinged stick, but wrongly identified the
center of percussion (at about two thirds of the height) as the probable
point of rupture. Reynolds\cite{Reynolds} first identified the possible
causes of the breaking with the shear forces and the bending moment
originating within the structure of the toppling chimney. More detailed
analyses were given by Bundy\cite{Bundy} and Madsen\cite{Madsen} (the most
complete papers we found on the subject) while simplified explanations are
also reported.\cite{Jones,Bartlett2,Walker} It even appears in graduate
student study guides,\cite{Cahn,Cronin} although the chimney is shown
bending the wrong way in one of these books.

In this paper we review the theory of the real falling chimney, outlined by
Madsen,\cite{Madsen} aiming for a complete and clear explanation of this
phenomenon in Sects. \ref{sect1} and \ref{sect2}. Then, in Sect. \ref{sect3}%
, we propose simple ways of using small scale models (literally \textit{toy
models }- made with toy blocks and bricks) to test effectively the outlined
theory. More information on these toy models can also be found on our
web-site,\cite{kaoru} together with photos and movie clips of the
experiments we have performed.

\section{Rotational motion of the falling chimney}

\label{sect1}

\begin{figure}[tbp]
\includegraphics{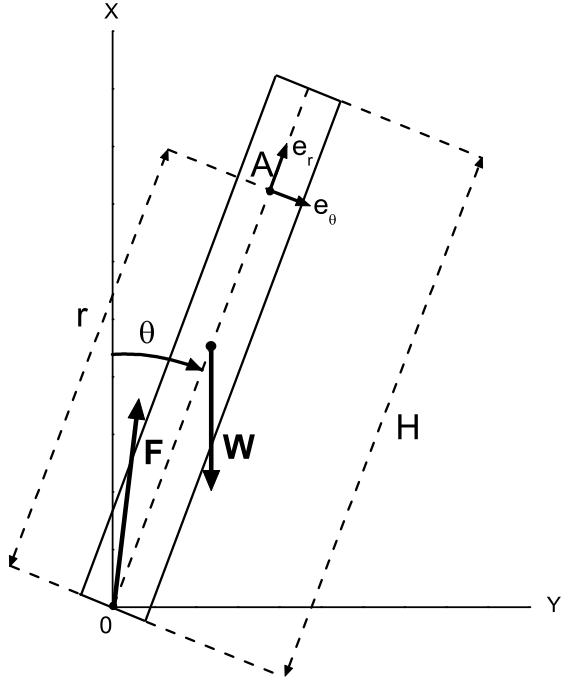}
\caption{The falling chimney described as a rotating uniform stick. The
external forces are the weight of the body applied to the center of gravity,
and the constraint force at the base.}
\label{fig1}
\end{figure}

The rotational motion of a falling chimney under gravity is equivalent to
that of the falling hinged stick of Sect. \ref{intro}. We can describe it as
in Fig. \ref{fig1}, 
where we use polar coordinates $r$ and $\theta $ (with $\widehat{e}_{r}$ and 
$\widehat{e}_{\theta }$ as unit vectors) for the position of an arbitrary
point $A$ on the longitudinal axis of the chimney, measuring the angle $%
\theta $ from the vertical direction. We treat the chimney as a uniform
rigid body of mass $m$ and height $H$, under the action of its weight $%
\mathbf{W}=m\mathbf{g}$, applied to the center of gravity (basically the
center of mass -CM- of the body), and a force $\mathbf{F}$ exerted by the
ground on the base of the chimney, assumed to act at a single point (we
neglect air resistance, or any other applied force). In plane polar
coordinates: 
\begin{subequations}
\begin{align}
\mathbf{W}& =W_{r}\widehat{e}_{r}+W_{\theta }\widehat{e}_{\theta }=-mg\cos
\theta \widehat{e}_{r}+mg\sin \theta \widehat{e}_{\theta }  \label{eqn1a} \\
\mathbf{F}& =F_{r}\widehat{e}_{r}+F_{\theta }\widehat{e}_{\theta }.
\label{eqn1b}
\end{align}

The moment of inertia of the chimney can be approximated with the one for a
uniform thin rod, with rotation axis perpendicular to the length and passing
through one end:\footnote{%
This is in general a good approximation. For a right circular hollow
cylinder of outer and inner radii $r_{1}$, $r_{2}$ and length $H$, hinged at
one end, the moment of inertia is $I=m\left( \frac{r_{1}^{2}+r_{2}^{2}}{4}+%
\frac{1}{3}H^{2}\right) \simeq \frac{1}{3}mH^{2}$ for $r_{1},r_{2}\ll H$.
For a typical chimney $r_{1}\lesssim \frac{H}{10}$, thus $\frac{r_{1}^{2}/4}{%
H^{2}/3}\lesssim 0.0075$, i.e., a correction of less than one percent. The
correction due to $r_{2}$ is even smaller.} 
\end{subequations}
\begin{equation}
I=\frac{1}{3}mH^{2}.  \label{eqn2a}
\end{equation}%
Applying the torque equation $I\overset{..}{\theta }=\tau _{z}$, for a
rotation around the origin, with an external torque given by $\tau _{z}=mg%
\frac{H}{2}\sin \theta $, we find the angular acceleration%
\begin{equation}
\overset{..}{\theta }=\frac{\tau _{z}}{I}=\frac{3}{2}\frac{g}{H}\sin \theta .
\label{eqn3}
\end{equation}%
A simple integration, using $\overset{..}{\theta }=\frac{d\overset{.}{\theta 
}}{dt}=\frac{d\overset{.}{\theta }}{d\theta }\overset{.}{\theta }=\frac{3}{2}%
\frac{g}{H}\sin \theta $, gives the angular velocity%
\begin{equation}
\overset{.}{\theta }^{2}=3\frac{g}{H}\left( 1-\cos \theta \right) ,
\label{eqn4}
\end{equation}%
assuming that the chimney starts moving from rest and is initially in the
vertical direction. A further integration of Eq. \ref{eqn4}, can lead to $%
\theta (t)$ in terms of elliptic integrals.

We recall that the acceleration in polar coordinates can be written as $%
\mathbf{a}$=$\overset{..}{\mathbf{r}}$=$(\overset{..}{r}-r\overset{.}{\theta 
}^{2})\widehat{e}_{r}+(r\overset{..}{\theta }+2\overset{.}{r}\overset{.}{%
\theta })\widehat{e}_{\theta }$, so that, for a point $A$ at a fixed
distance $r$ from the origin, it becomes%
\begin{equation}
\mathbf{a}=a_{r}\widehat{e}_{r}+a_{\theta }\widehat{e}_{\theta }=-r\overset{.%
}{\theta }^{2}\widehat{e}_{r}+r\overset{..}{\theta }\widehat{e}_{\theta }.
\label{eqn5}
\end{equation}%
For a point at two thirds of the height, $r=\frac{2}{3}H$, combining Eqs. %
\ref{eqn3} and \ref{eqn5} we get $a_{\theta }(r=\frac{2}{3}H)=\frac{2}{3}H%
\overset{..}{\theta }=g\sin \theta $, proving that this particular point is
the \textit{center of percussion} of the body, as already mentioned in Sect. %
\ref{intro}.

The torque equation allowed us to determine the angular acceleration of the
motion in Eq. \ref{eqn3}. We can use this result and Newton's second law for
the motion of the center of mass (CM) of the whole chimney to determine the
unknown force $\mathbf{F}$ at the base. The vectorial equation is%
\begin{equation}
m\overset{..}{\mathbf{r}}_{CM}=\mathbf{W}+\mathbf{F}  \label{eqn6}
\end{equation}%
which, for $r=\frac{H}{2}$, splits into radial and angular equations, 
\begin{subequations}
\begin{align}
-m\frac{H}{2}\overset{.}{\theta }^{2}& =F_{r}-mg\cos \theta  \label{eqn7a} \\
m\frac{H}{2}\overset{..}{\theta }& =F_{\theta }+mg\sin \theta .
\label{eqn7b}
\end{align}%
Using Eqs. \ref{eqn3} and \ref{eqn4}, the two components of the force $%
\mathbf{F}$ are easily determined: 
\end{subequations}
\begin{subequations}
\begin{align}
F_{r}& =\frac{5}{2}mg\left( \cos \theta -\frac{3}{5}\right)  \label{eqn8a} \\
F_{\theta }& =-\frac{1}{4}mg\sin \theta .  \label{eqn8b}
\end{align}

\section{Internal forces and bending moment}

\label{sect2}

We now move to the analysis of the internal forces which develop inside the
structure of the falling chimney. The resulting stresses and bending moment
are the causes of the rupture of the toppling chimney. Consider, as in Fig. %
\ref{fig2}, 
\begin{figure}[tbp]
\includegraphics{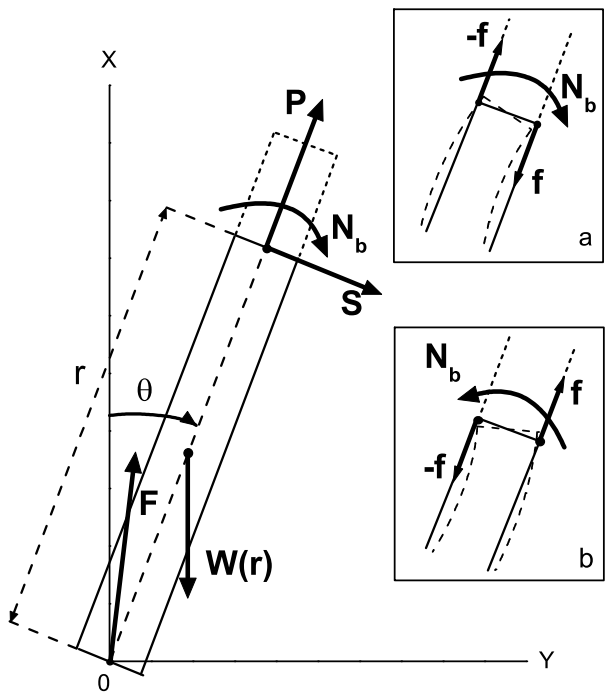}
\caption{The forces acting on the lower portion of the chimney, due to the
upper part and the action of the constraint at the base, are shown here.
The two insets explain the definition of the bending moment in terms of a 
couple of forces. The resulting deformation of the structure is also 
shown for the two possible cases.}
\label{fig2}
\end{figure}
an arbitrary lower portion of the chimney of height $r$ (as opposed to the
total height $H$) and the forces acting on this part of the structure due to
the upper portion and the base. The weight of the lower portion is now $%
W(r)=mg\frac{r}{H}$ (assuming again a uniform structure, so that the weight
is proportional to the height of the considered portion) and it is applied
to the center of gravity of this lower portion (at a distance $\frac{r}{2}$
from the origin). In polar coordinates: 
\end{subequations}
\begin{equation}
\mathbf{W}(r)=W_{r}(r)\widehat{e}_{r}+W_{\theta }(r)\widehat{e}_{\theta }=-mg%
\frac{r}{H}\cos \theta \widehat{e}_{r}+mg\frac{r}{H}\sin \theta \widehat{e}%
_{\theta }.  \label{eqn9}
\end{equation}

The force $\mathbf{F}$, applied at the base, is still the same as in Eqs. %
\ref{eqn8a}, \ref{eqn8b}, but we have to add the action of the upper part on
the lower portion. We follow here the general analysis of the internal
forces and moments which can be found in every textbook on Statics (see for
example \cite{Statics,Statics2}) and which can be easily adapted to our case.

The distribution of all the internal forces, at the cross section
being considered, can be equivalently described by a resultant force and a
resultant moment acting at a specific point of the cross section (typically
the ``centroid'' of the sectioned area, in our case simply the central point
of the section, on the longitudinal axis). In particular, the resultant
force can be decomposed into a transverse shearing force $\mathbf{S}%
=S_{\theta }\widehat{e}_{\theta }$, and a longitudinal stress force (tension
or compression) $\mathbf{P}=P_{r}\widehat{e}_{r}$, applied as in Fig. \ref%
{fig2}, at the cross section between the upper and lower portions, and
assumed positive in the $\widehat{e}_{\theta }$, $\widehat{e}_{r}$ direction
respectively.

In addition, we have to consider the resultant moment of the forces at the 
cross section, which is 
usually called the ``bending moment'' $\mathbf{N}_{b}$, because its effect
will ultimately result in bending the structure. It is represented in the
picture by the curved arrow. Since we treat this as a plane problem, the
bending moment can only have a component perpendicular to the plane of the
figure, i.e., in the $z$ direction. No other components are considered here,
in particular we assume that no torsional moment exist in the structure,
which would tend to twist the chimney around its longitudinal axis.

The bending moment $\mathbf{N}_{b}=N_{b}\widehat{e}_{z}$ can be thought as
originating from a couple of forces, $\mathbf{f}$ and $-\mathbf{f}$, that
can be regarded as applied to the leading and trailing edge of the
structure, at the cross section considered. This couple of forces is shown
explicitly in the papers by Bundy\cite{Bundy} and Madsen,\cite{Madsen} but
we prefer to use directly $\mathbf{N}_{b}$ in our
treatment, because the bending moment is the result of the whole distribution
of forces at the cross section considered. The two small insets inside Fig. %
\ref{fig2} explain the definition of the bending moment in terms of the
couple of forces $\mathbf{f}$ and $-\mathbf{f}$. We also show the resulting
deformation of the structure due to a ``clockwise'' (diagram a), or a
``counter-clockwise'' bending moment (diagram b). The latter case will be
the actual deformation of the falling chimney. $N_{b}$ will be assumed to be
positive if it acts as in the figure, i.e., a positive component of the
torque in the $z$ direction (we assume here the use of a right-handed system
of coordinate axis). In the following we will refer to $N_{b}$ as the
bending moment, acting on the lower portion of the chimney.\footnote{%
The bending moment and the stress forces can be thought as applied to the
center (centroid) of the cross sectional area we are considering, on the
face belonging to the lower portion of the structure. Equal and opposite
forces and moments would originate on the face belonging to the upper
portion.}

Again, we will consider the torque equation and the second law for the
motion of the center of mass (located at $\frac{r}{2}$) for just the lower
portion of the chimney (of mass $\frac{r}{H}m$). It is better to analyze the
CM motion first. The vector equation%
\begin{equation}
m\frac{r}{H}\overset{..}{\mathbf{r}}_{CM}=\mathbf{W}(r)+\mathbf{F}+\mathbf{P}%
+\mathbf{S,}  \label{eqn10}
\end{equation}%
will split into the radial and angular directions, 
\begin{subequations}
\begin{eqnarray}
-m\frac{r^{2}}{2H}\overset{.}{\theta }^{2} &=&-\frac{3}{2}mg(1-\cos \theta )%
\frac{r^{2}}{H^{2}}=-mg\frac{r}{H}\cos \theta +\frac{5}{2}mg\left( \cos
\theta -\frac{3}{5}\right) +P_{r}  \label{eqn11a} \\
m\frac{r^{2}}{2H}\overset{..}{\theta } &=&\frac{3}{4}mg\sin \theta \frac{%
r^{2}}{H^{2}}=mg\frac{r}{H}\sin \theta -\frac{1}{4}mg\sin \theta +S_{\theta
},  \label{eqn11b}
\end{eqnarray}%
having used Eqs. \ref{eqn3}, \ref{eqn4}, \ref{eqn8a}, \ref{eqn8b}, and \ref%
{eqn9}. We can solve for the longitudinal and transverse forces 
\end{subequations}
\begin{subequations}
\begin{eqnarray}
P_{r} &=&-\frac{1}{2}mg\left( 1-\frac{r}{H}\right) \left[ \left( 5+3\frac{r}{%
H}\right) \cos \theta -3\left( 1+\frac{r}{H}\right) \right]  \label{eqn12a}
\\
S_{\theta } &=&\frac{3}{4}mg\sin \theta \left( \frac{r^{2}}{H^{2}}-\frac{4}{3%
}\frac{r}{H}+\frac{1}{3}\right) ,  \label{eqn12b}
\end{eqnarray}%
which depend on the fraction of height $\frac{r}{H}$, the angle of rotation $%
\theta $, and also the total weight $mg$. Following the analysis by Bundy,%
\cite{Bundy} we plot these two forces in Figs. 3 and 4 respectively,
normalized to the total weight $mg$, as a function of the height fraction,
for several angles.


\begin{figure}[tbp]
\includegraphics{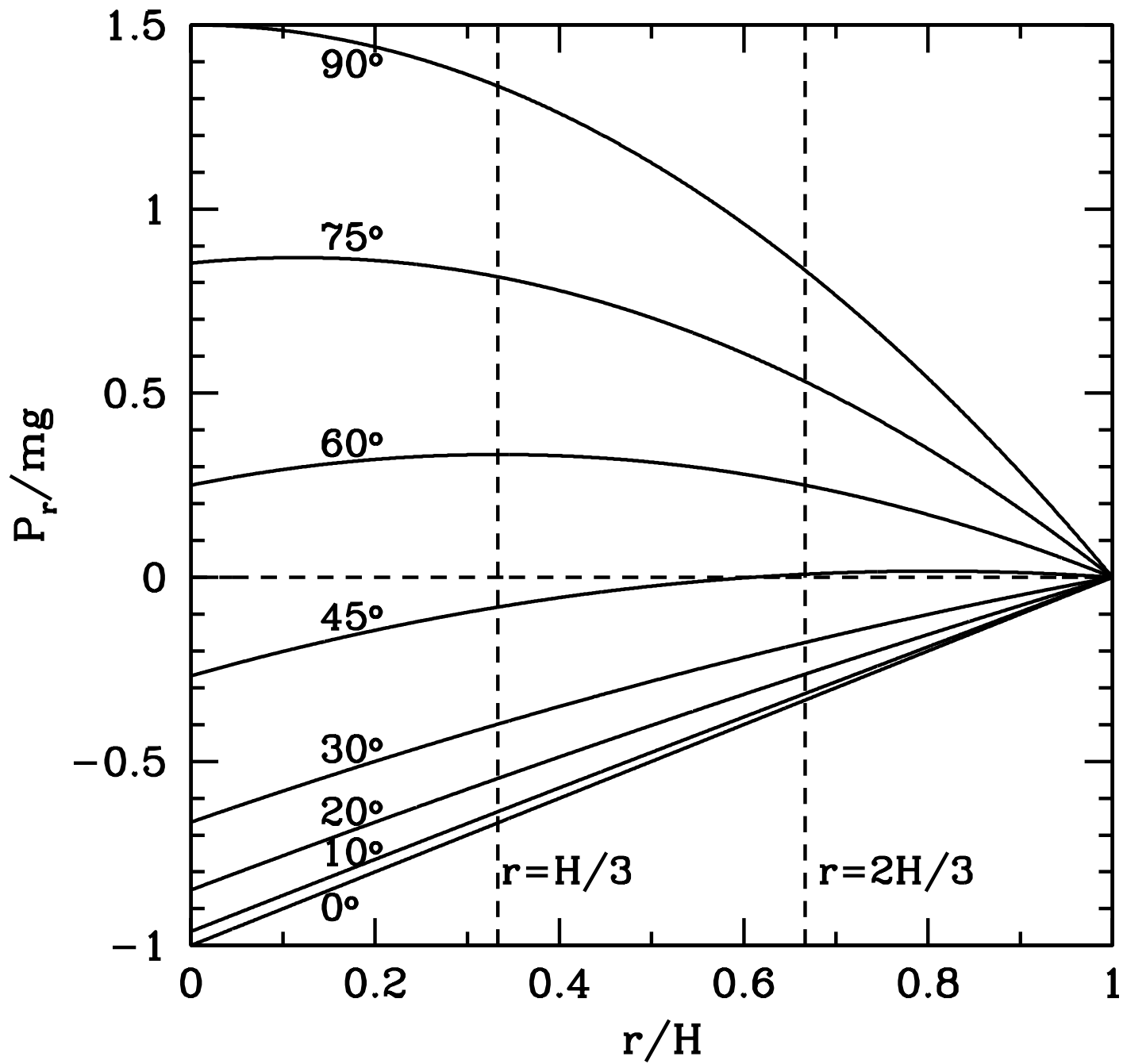}
\caption{The longitudinal stress force per unit weight of the chimney is
shown as a function of the height fraction for several angles. Positive
values indicate tensions, while negative values represent compressions.}
\label{fig3}
\end{figure}

From Fig. \ref{fig3} we see that $P_{r}$ is negative (a compression) for
smaller angles, but eventually becomes positive (a tension) for angles
greater than about $45^{\circ }$. $P_{r}$ also depends critically on $\frac{r%
}{H}$ (for $\theta =0^{\circ }$, $P_{r}$ represents simply the compression
due to the weight of the upper part acting on the lower part). This
longitudinal force will be combined later with the bending moment to
determine the total stress at the leading and trailing edges, which is the
most typical cause of the rupture.


\begin{figure}[tbp]
\includegraphics{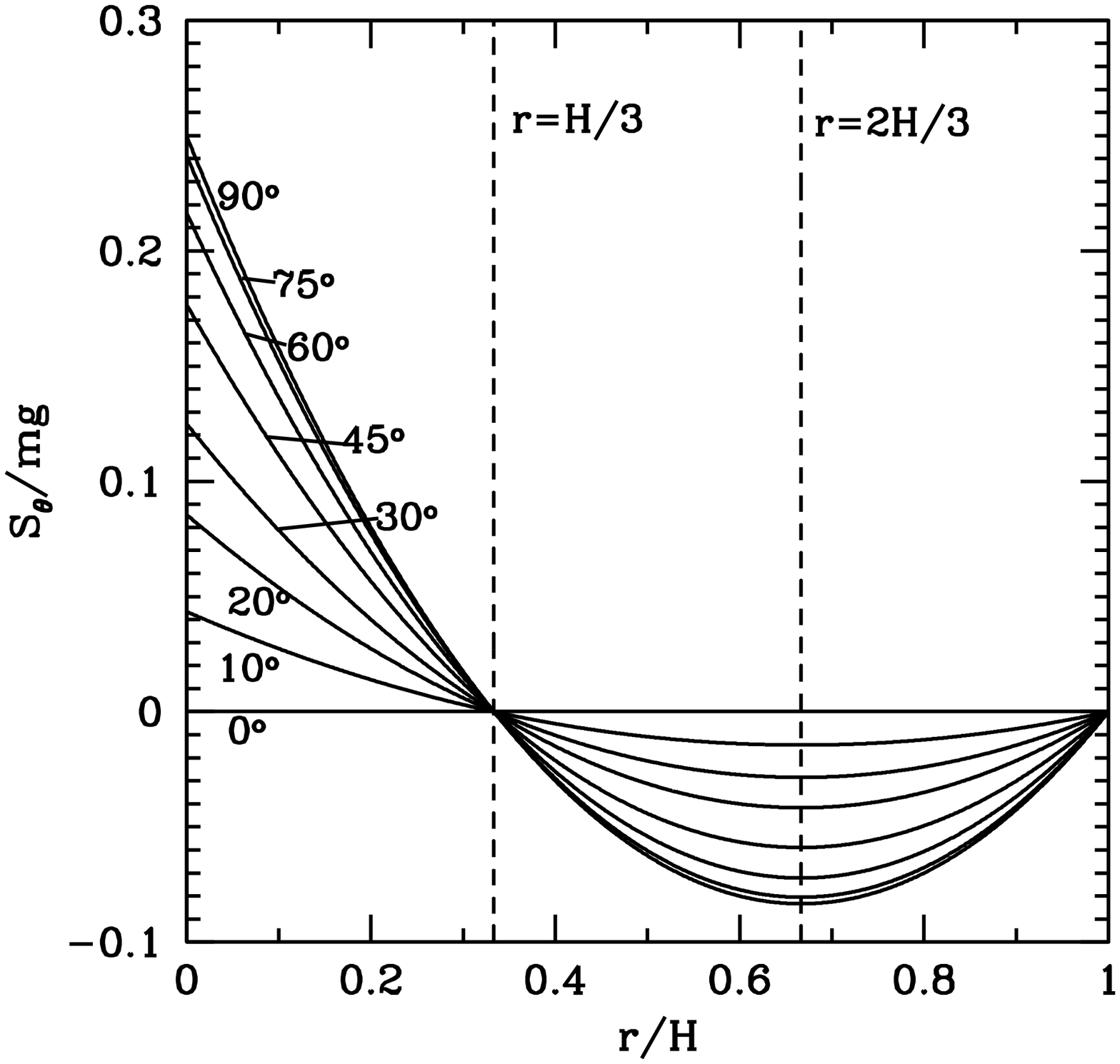}
\caption{The transverse shear force per unit weight of the chimney is shown
as a function of the height fraction for several angles. Positive values are
for forces in the $\widehat{e}_{\protect\theta }$ direction.}
\label{fig4}
\end{figure}

In Fig. \ref{fig4} we plot the (transverse) shear force $S_{\theta }$, which
can be the other leading cause of rupture. It is easily seen that, for any
considered angle, the magnitude of the shear force $\left| S_{\theta
}\right| $ has an absolute maximum at $\frac{r}{H}=0$ (and a positive
value), meaning that large shear forces, in the $\widehat{e}_{\theta }$
direction, usually originate near the base. The shear force is always zero
at one third of the height, and $\left| S_{\theta }\right| $ also has a
(relative) maximum at $\frac{2}{3}H$ (with a negative value, therefore $%
S_{\theta }$ is in the $-\widehat{e}_{\theta }$ direction), but this value
is smaller than the one near the base.

From this analysis, it is typically concluded that if the structure breaks
from shear stress alone, this is usually more likely to happen near the
base. This can be seen for example in the already mentioned cover photo of
the September 1976 issue of The Physics Teacher,\cite{Bartlett2,kaoru}
showing the fall of a chimney in Detroit. The two ruptures at the bottom are
likely due to shear forces, while a third rupture can be seen at about $%
r=0.47H$, and this is due to the combination of bending moment and
longitudinal force $P_{r}$, as we will explain in the following. More photos
and detailed pictorial descriptions of chimney ruptures can be found in the
paper by Bundy.\cite{Bundy,kaoru}

The ``bending moment'' $N_{b}$ can be calculated from the torque equation $%
I(r)\overset{..}{\theta }=\tau _{z}$, where now $I(r)=\frac{1}{3}m\frac{r}{H}%
r^{2}$ is the moment of inertia of just the lower part. $\overset{..}{\theta 
}$ will come from Eq. \ref{eqn3}, and the total external torque is now $\tau
_{z}=\frac{r}{2}W_{\theta }(r)+rS_{\theta }+N_{b}$. Using Eqs. \ref{eqn9}
and \ref{eqn12b}, we can solve the torque equation for the bending moment,
obtaining 
\end{subequations}
\begin{equation}
N_{b}=-\frac{1}{4}mg\sin \theta \ r\left( 1-\frac{r}{H}\right) ^{2},
\label{eqn13}
\end{equation}%
or, in a non-dimensional form 
\begin{equation}
\frac{N_{b}}{mgH}=-\frac{1}{4}\sin \theta \ \frac{r}{H}\left( 1-\frac{r}{H}%
\right) ^{2},
\label{eqn13bis}
\end{equation}
\begin{figure}[tbp]
\includegraphics{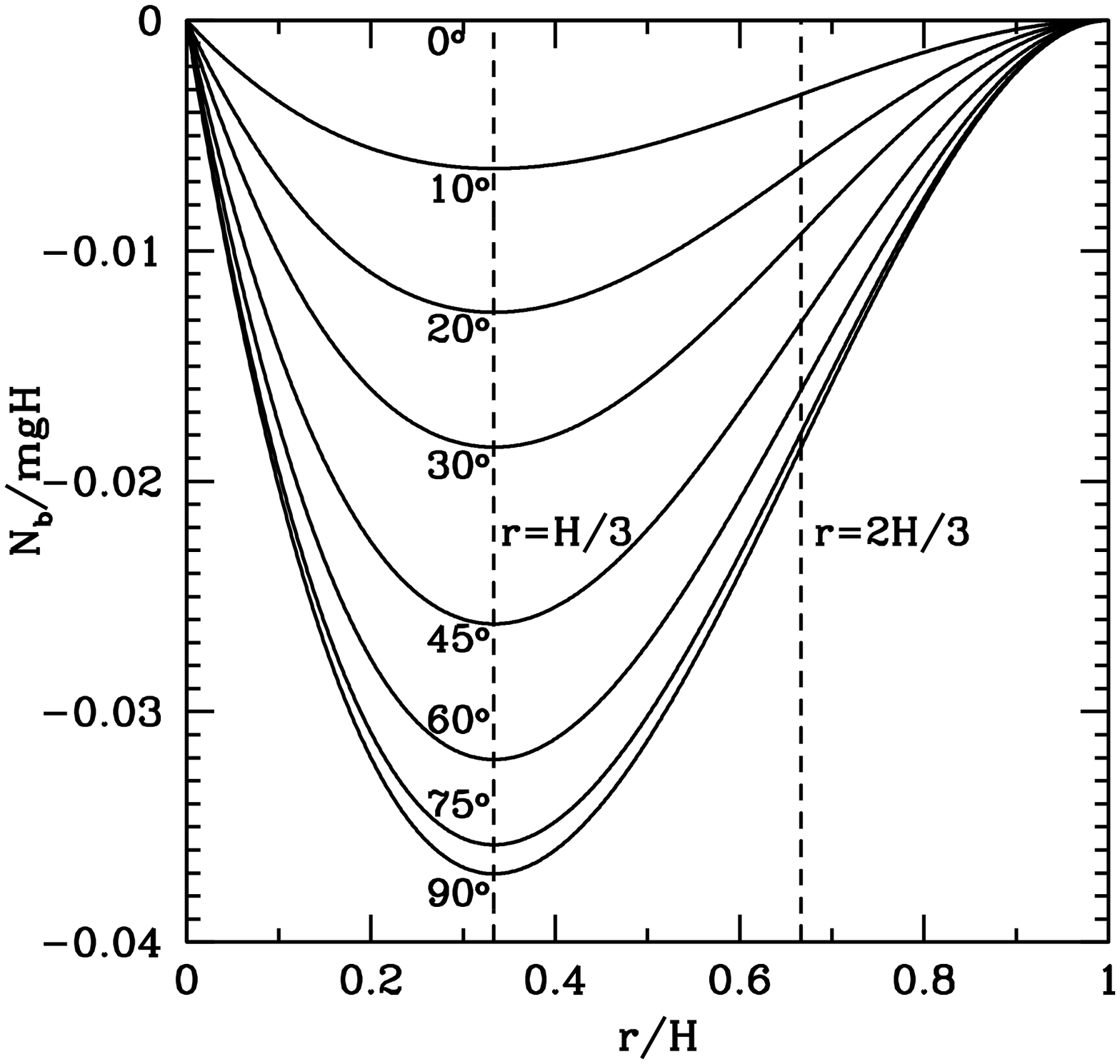}
\caption{The bending moment, divided by the weight and the height of the
chimney, is shown as a function of the height fraction for several angles.}
\label{fig5}
\end{figure}
which is plotted in Fig. 5, as a function of the height fraction and for
various angles.

$N_{b}$ is always negative, showing that it is actually directed in the
opposite way of Fig. \ref{fig2} (or as in diagram b of Fig. \ref{fig2}). This
particular direction of the bending moment will induce a tension in the
leading edge of the chimney and a compression in the trailing edge. The
structure will bend accordingly, with the concavity on the side of the
trailing edge, and will eventually break in the way shown by the many
existing photos. We can see that, for any angle, the bending moment is
obviously zero at the bottom and at the top of the chimney, while it assumes
its maximum absolute value at exactly one third of the height $H$. The
bending moment alone would therefore induce a rupture at one third of the
structure, but the total longitudinal stress at the leading edge is also due
to the force $P_{r}$, as we will show next.

Another interesting relation between the bending moment $N_{b}$ and the
shear force $S_{\theta }$ is that they are in general related by a simple
derivative, i.e., $S_{\theta }=-\frac{\partial N_{b}}{\partial r}$, as it is
easy to check from Eq. \ref{eqn12b} and Eq. \ref{eqn13}.\footnote{%
A positive sign in the relation between bending moment and shear is usually
reported in textbooks on Statics, due to a different choice of the sign of
the moment.} This is a well known relationship of the Statics of beams and
other structural members (for a complete proof see for example Hibbeler\cite%
{Statics2}). It is a direct consequence of the equilibrium equations applied
to an infinitesimal longitudinal portion of the beam: the change in bending
moment along the beam is always equal to the shear force applied to that
portion of the beam.

Finally, we can combine $N_{b}$ and $P_{r}$ to compute the total
longitudinal stress on the cross sectional area between the lower and upper
parts. We follow the theory of elasticity and deformations in beams, which
can be found in classical treatises such as Sommerfeld\cite{Sommerfeld} and
Landau-Lifshitz,\cite{Landau} or again in Bundy's paper.\cite{Bundy} The
longitudinal stress is maximum at the leading and trailing edges, located at
the maximum distance from the longitudinal (centroidal) axis of the chimney
which lies within the ``neutral surface'' of the structure, the surface
which is neither stretched nor compressed.

\begin{figure}[tbp]
\includegraphics{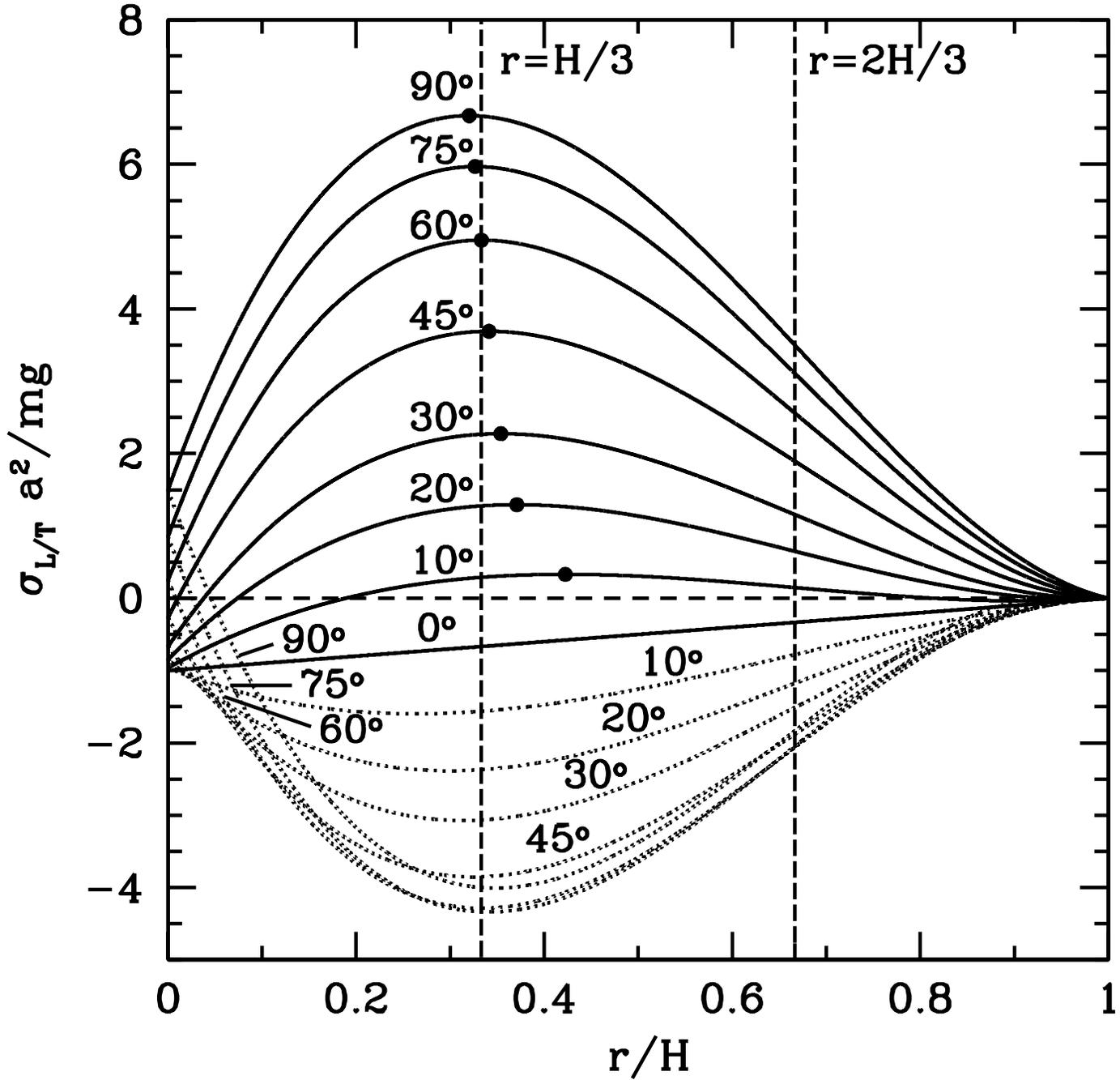}
\caption{The normalized longitudinal stress at the leading edge (solid) and
at the trailing edge (dotted), are shown as a function of the height
fraction and for several angles ($H/a=24$). The maxima of the leading edge
stress curves are marked by solid points.}
\label{fig6}
\end{figure}

For simplicity, we will only consider from here on, structures with uniform
square cross section of side $a$, as this is the case of the toy models
described in Sect. \ref{sect3}. In this case the stresses at the leading and
trailing edge, $\sigma _{L}$ and $\sigma _{T}$ respectively, can be
evaluated from 
\begin{equation}
\sigma _{L/T}=\frac{P_{r}}{A}\mp \frac{aN_{b}}{2J}  \label{eqn14}
\end{equation}%
(the upper sign is for $\sigma _{L}$, the lower for $\sigma _{T}$) where $%
A=a^{2}$ is the area of the square cross section of side $a,$ with the
factor $\frac{a}{2}$ representing the distance between the longitudinal
axis, considered as the ``neutral axis,'' and the two edges. $J=\frac{a^{4}}{%
12}$ is the moment of inertia of the cross sectional area computed about the
neutral axis (see Sommerfeld\cite{Sommerfeld} for details). Using also the
expressions for $P_{r}$ and $N_{b}$, we obtain%
\begin{equation}
\frac{\sigma _{L/T}\ a^{2}}{mg}=-\frac{1}{2}\left( 1-\frac{r}{H}\right) %
\left[ \left( 5+3\frac{r}{H}\right) \cos \theta -3\left( 1+\frac{r}{H}%
\right) \right] \pm \frac{3}{2}\frac{H}{a}\sin \theta \frac{r}{H}\left( 1-%
\frac{r}{H}\right) ^{2},  \label{eqn15}
\end{equation}%
where we normalized $\sigma _{L/T}$, dividing by $\frac{mg}{a^{2}}$, in
order to obtain a dimensionless quantity which is plotted in Fig. \ref{fig6}%
, as a function of the height fraction and for several angles.


This quantity depends also on the ratio $\frac{H}{a}$, which for a real
chimney is of the order $\frac{H}{a}\gtrsim 10$. For the toy models
described in Sect. \ref{sect3}, the value of this ratio is even bigger: $%
\frac{H}{a}\simeq 24-61$, enhancing the contribution of the second term of
Eq. \ref{eqn15}, which comes from the bending moment $N_{b}$. In Fig. \ref%
{fig6} we show the plot for $\frac{H}{a}=24$, but similar figures can be
obtained for different values of the ratio.

In Eqs. \ref{eqn14}-\ref{eqn15}, and in Fig. \ref{fig6} the total stresses
are considered positive if they represent tensions, negative if they are
compressions. It is easily seen from the figure that the stress at the
leading edge $\sigma _{L}$, is initially a compression, but eventually
becomes a tension, constantly increasing for larger angles; $\sigma _{T}$ on
the contrary is usually a compression. It is therefore the combination of
these intense tensile stresses in the leading part of the chimney (and also
compressions on the trailing side) that causes the rupture of the chimney.

This type of breaking is more likely to occur at the positive maximum value
of $\sigma _{L}$. This maximum value depends critically on the $\frac{r}{H}$
ratio, for a certain angle of rupture $\theta $, so it is possible, just by
looking at the maxima of the solid curves in Fig. \ref{fig6} (marked by
solid points), to roughly match the height of the rupture point to the angle
at which the breaking started to occur. As far as the actual prediction of
the point of rupture, this would obviously depend on the building materials
and the construction of the chimney or tower, an analysis of which goes
beyond the scope of this work.

It is interesting to note, from Fig. \ref{fig6} again, that the stress $%
\sigma _{L}$ is not always maximum at one third of the height (as for the
bending moment of Fig. \ref{fig5}). For small angles of about $\theta \simeq
5^{\circ }-20^{\circ }$ it reaches a maximum for $\frac{r}{H}\simeq 0.4-0.5$%
, while for larger angles it approaches the typical ratio $\frac{r}{H}\simeq 
\frac{1}{3}$. This means that if the chimney breaks early in its fall, for
small angles, it is more likely to break near the center; on the contrary if
the rupture occurs at larger angles, the breaking point is usually shifted
toward one third of the height. This is the effect of the $P_{r}$ term in
Eq. \ref{eqn14}, which modifies the position of the maximum $\sigma _{L}$
(or $\sigma _{T}$).

\begin{figure}[tbp]
\includegraphics{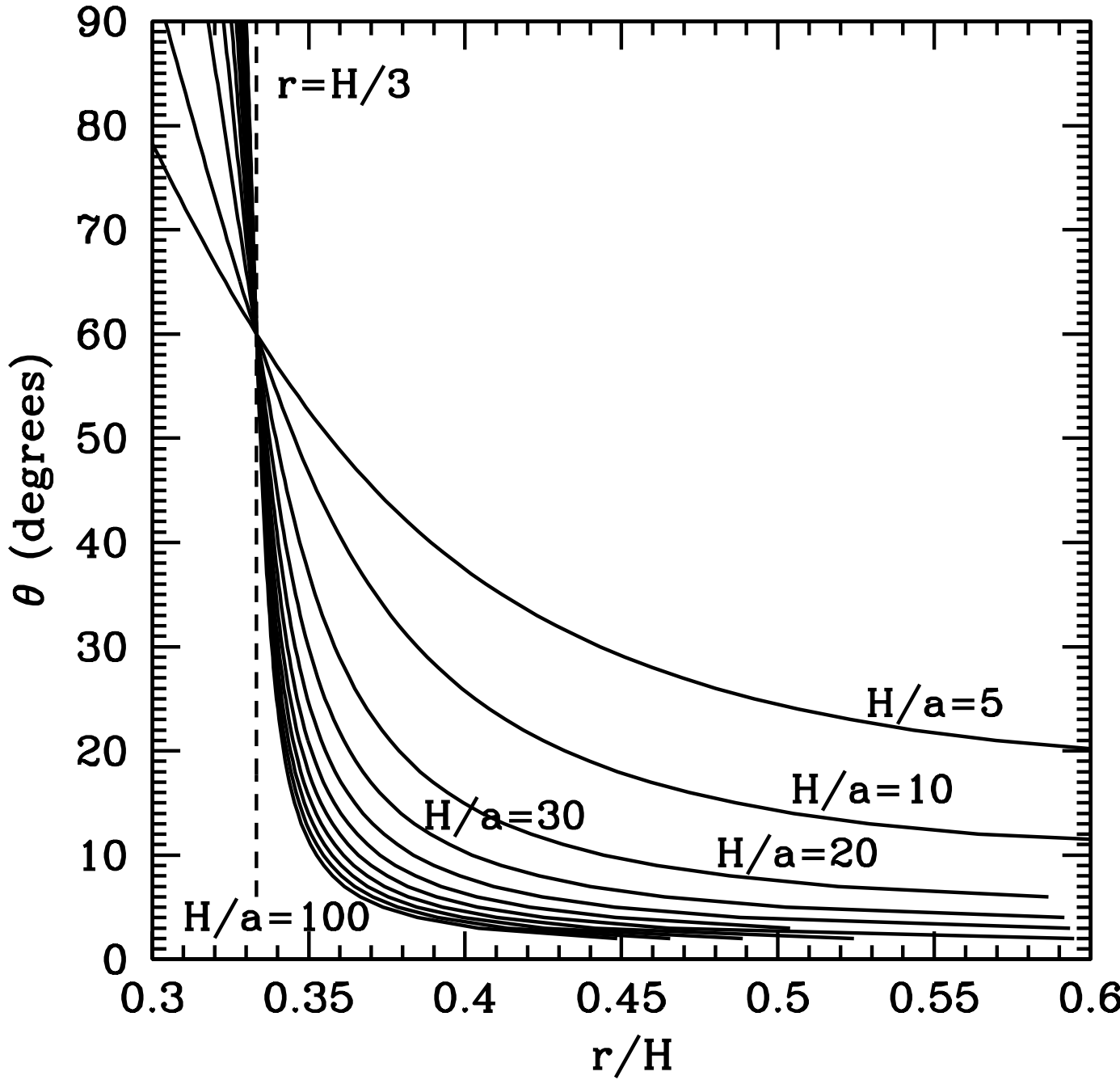}
\caption{The maxima of the leading edge stress curves are shown as continuous functions, for several values of the ratio $H/a$. They represent the points, in terms of the height ratio $r/H$ and angle $\theta$, at which the structure is more likely to break, due to bending.}
\label{fig6bis}
\end{figure}

As already mentioned, another factor to be considered is the ratio $\frac{H}{%
a}$, which is usually in the range $\frac{H}{a}\simeq 5-20$ for real
chimneys, but can be increased up to about $\frac{H}{a}\simeq 100$ with our
toy models. In Fig. \ref{fig6bis} we plot the maxima of the leading edge
stress curves, for several values of the ratio $\frac{H}{a}%
=5,10,20,30,...,100$ (for the group of curves between the values of $30$ and 
$100$, the parameter is increased by $10$ units for each curve).

These maxima are plotted as continuous curves, showing the corresponding
values of the ratio $\frac{r}{H}$ and the angle $\theta $, at which the
structure is more likely to break, for a given value of $\frac{H}{a}$. In
other words, connecting for example the solid points of Fig. \ref{fig6}, we
would obtain a corresponding continuous curve in Fig. \ref{fig6bis}, with
the angle of probable rupture on the vertical axis, instead of the
normalized stress.

The dependence of the rupture point on these quantities can be noted in
several of the existing photos of falling chimneys, when the breaking occurs
due to the bending of the structure and not for the transverse shear stress
near the base mentioned at the beginning of this section. The angle of
rupture can be roughly estimated by measuring the angle the upper part of
the chimney forms with the vertical in the photos. This angle tends in fact
not to change much after the rupture, since the upper part falls without
much additional rotation. For example the photos in the paper by Bartlett%
\cite{Bartlett2,kaoru} refer to chimneys with $\frac{H}{a}\simeq 10$,\
breaking at about $\theta \simeq 20^{\circ }-25^{\circ }$ and for $\frac{r}{H%
}\simeq 0.47$, consistent with the values of Fig. \ref{fig6bis}, for the $%
\frac{H}{a}=10$ curve. Similar behavior can be seen in other photos,\cite%
{Bundy,kaoru} since real chimneys tend to break very early in their fall,
due to the intense stresses originating within their structure.

\section{Toy models}

\label{sect3}

In this section we discuss our efforts to reproduce the effects described
above, with the help of toy models of the falling chimney. These models were
constructed with simple toy blocks of different type and size, and their
fall was filmed with a digital camera, so that we could analyze the events
frame by frame, to test the theory. Complete details on the type of blocks
used, experimental settings and video-capture techniques, as well as the
complete set of our video-recordings and still photos can be found on our
web-site,\cite{kaoru} and in an upcoming publication.\cite{kaoru2}

Bundy noted in his work\cite{Bundy} that the use of a model to test a real
chimney would be useless due to a ``scale effect.'' The stresses inside the
chimney depend roughly on the scale of the object, so that real chimneys
would develop bigger stresses than equivalent small-scale models, therefore
breaking earlier in their fall. Nevertheless we found it interesting to
reproduce these effects in small scale models to test especially the
discussion based on Figs. \ref{fig6} and \ref{fig6bis}. It was also supposed
to be difficult\cite{Bundy} to show these effects with toy models.\footnote{%
We did not find any use of toy models for this problem, either in the
literature or in web-pages devoted to physics demos. We will be grateful to
receive information about similar experiments, if any.}


\begin{figure}[tbp]
\includegraphics{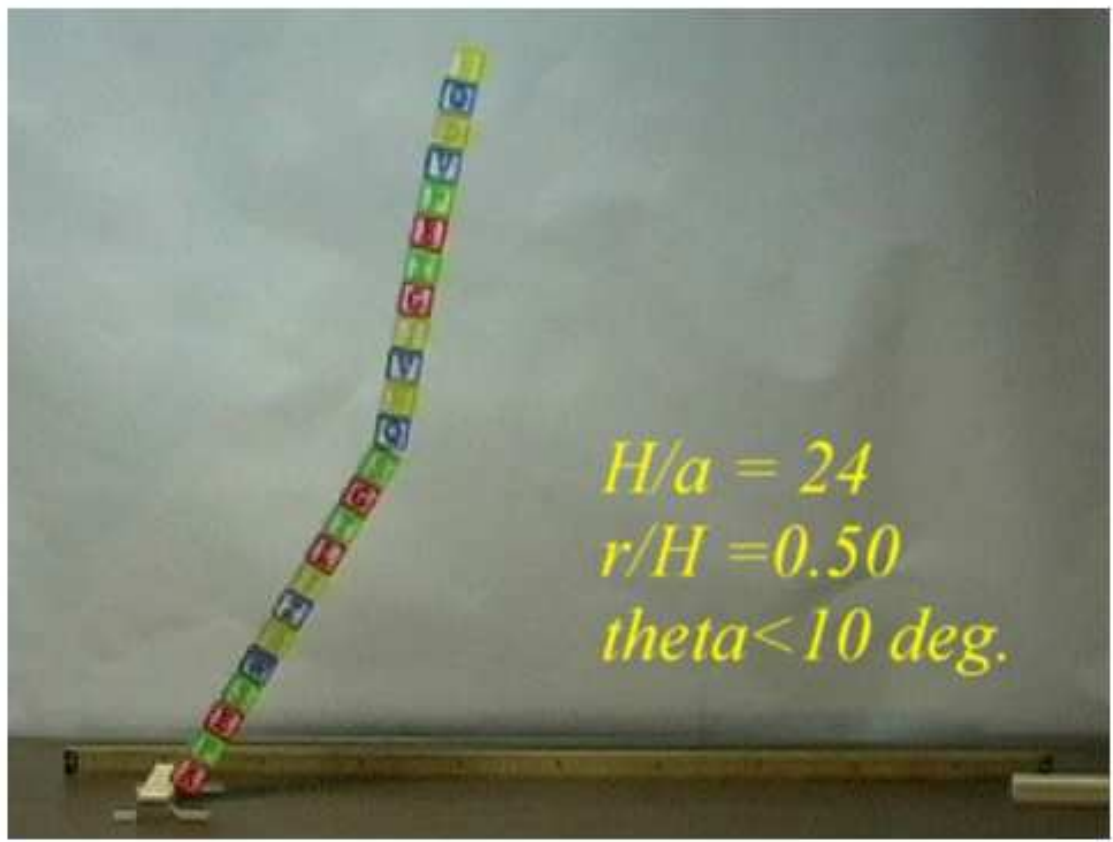}
\caption{The first toy model made with wooden blocks. The structure appears
to break at $\frac{r}{H}=0.5$, and at a small angle $\protect\theta \lesssim
10^{\circ }$, which can be estimated by measuring the angle formed by the
upper part of the tower with the vertical.}
\label{fig7}
\end{figure}

Fig. \ref{fig7} is the first example of one of our toy models. We made a
tower by simply stacking $24$ wooden toy blocks of cubic shape, for a total
height $H=0.76\ m$, mass $m=0.32\ kg$ and a ratio $\frac{H}{a}=24$, the
value used in Fig. \ref{fig6}. The tower was set into the falling motion by
removing a support at the bottom, inducing a rotation without slipping at
the bottom point. The picture clearly shows the ``rupture'' due to bending
of the structure at exactly half the height, $\frac{r}{H}=0.50$, and for a
small angle $\theta \lesssim 10^{\circ }$, which again can be estimated by
measuring the angle the upper part of the chimney forms with the vertical
direction. This is in good agreement with the position of the maximum for
the solid-$10^{\circ }$ curve in Fig. \ref{fig6} and also with the data of
Fig. \ref{fig6bis}, using the $\frac{H}{a}=20$ curve, showing that the 
theory is applicable also to these small-scale models.


\begin{figure}[tbp]
\includegraphics{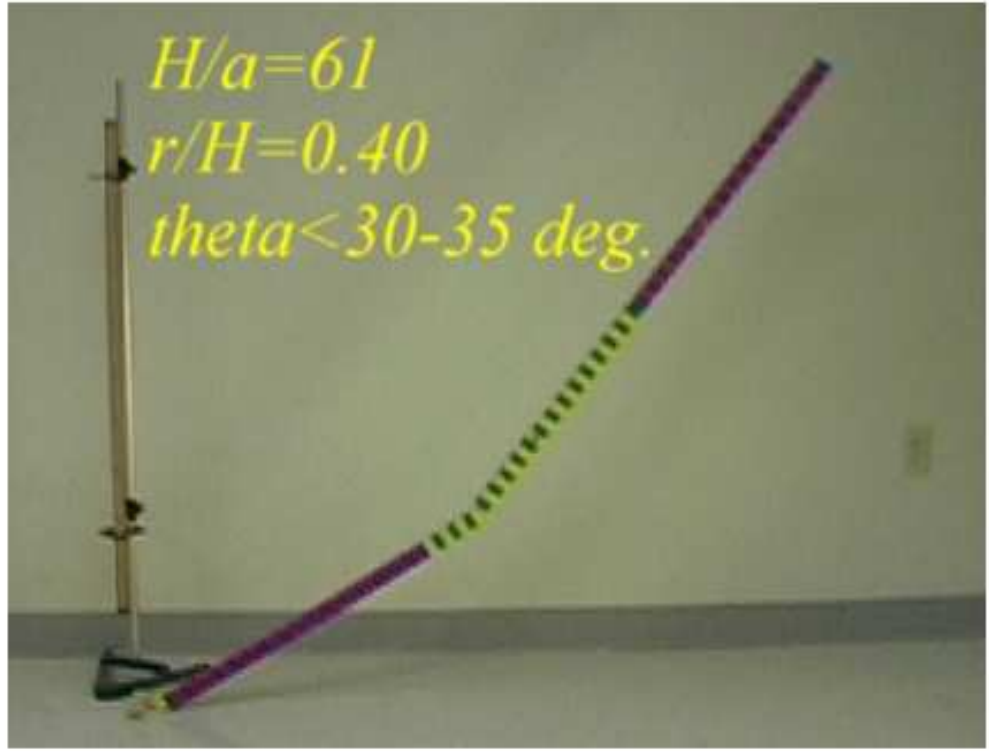}
\caption{The second toy model made with plastic blocks. The structure
appears to break at $\frac{r}{H}=0.40 $, and at a larger angle $\protect%
\theta \simeq 30^{\circ }-35^{\circ } $, which can be estimated by measuring
the angle formed by the upper part of the tower with the vertical.}
\label{fig8}
\end{figure}

Our second example is a taller tower ($H=1.9\ m$, $m=0.65\ kg$, and $\frac{H%
}{a}=61$) made with 100 plastic blocks of a very popular brand of toy
bricks. The blocks are inserted on top of each other so that bending of the
structure is allowed, but shear stress cannot possibly break the tower. The
100 toy blocks are arranged by color to subdivide the structure into three
equal parts, and also the position of the center is marked. This time the
rupture occurs for an angle around $\theta \simeq 30^{\circ }-35^{\circ }$,
and at the height ratio $\frac{r}{H}=0.40$. This is consistent with
the $30^{\circ }$ solid curve in Fig. \ref{fig6}, while the data from 
Fig. \ref{fig6bis} (for $\frac{H}{a}=60$) would suggest a smaller angle of 
rupture. It is actually difficult, with this type of toy bricks, to estimate 
the angle at which the structure begins to bend.

We performed several other experiments, varying the dimensions of the
towers, the type of blocks, always obtaining results consistent with the
theory. We can conclude that it is actually easy to reproduce the bending
and breaking of chimneys with small scale towers, and this type of
experiment could be made part of an undergraduate laboratory class for
rotational mechanics, with some minor adaptations and changes.\cite{kaoru2}

\section{Conclusion}

\label{conclusion}

In this paper we reviewed the theory of the falling chimney, showing that
the rupture can be caused by either shear forces typically near the base, or
by the bending of the structure which is caused primarily by the bending
moment, but is also affected by the longitudinal stress force. In the latter
case the breaking is more likely to occur between one third and one half of
the height of the chimney.

This point of rupture is also related to the angle of rupture and this
relationship can be verified in the many existing photographic reports of
falling chimneys. We also constructed several small scale toy models,
showing that it is possible with them to reproduce the dynamics of the fall.
By examining photos taken during the fall of these models we were able to
confirm the theoretical model outlined in this paper.

\begin{acknowledgments}
This research was supported by an award from Research Corporation. The
authors would like to thank Dr. V. Coletta for the many useful discussions
regarding this paper. G.V. would like to acknowledge and thank his friend
and physics teacher, Prof. G. Tonzig, whose excellent book\cite{Tonzig}
``Cento Errori di Fisica'' inspired the original idea for this paper.
\end{acknowledgments}


\bibliographystyle{utphys}
\bibliography{chimney}


\end{document}